\documentclass[journal]{new-aiaa} %for journal papers
\usepackage[utf8]{inputenc}
\usepackage{textcomp}
\usepackage{xcolor}

\usepackage{graphicx}
\usepackage{amsmath}
\usepackage[version=4]{mhchem}
\usepackage{siunitx}
\usepackage{longtable,tabularx}
\usepackage{subcaption}
\usepackage{bm}
\usepackage{breqn}
\usepackage{tikz}
\usepackage{doi}

\usetikzlibrary{shapes,arrows}

\title{Adaptive and Efficient Model Predictive Control for Booster Re-entry}

\author{Joseph Chai\footnote{PhD Candidate, School of Mechanical and Mining Engineering; j.chai@uq.edu.au}}
\affil{University of Queensland, Brisbane, QLD 4067, Australia}
\author{ Eran Medagoda\footnote{Guidance, Navigation, and Control Lead; e.medagoda@skybornetech.com}}
\affil{Skyborne Technologies, Brisbane, QLD 4172, Australia}
\author{ Erkan Kayacan\footnote{Lecturer, School of Mechanical and Mining Engineering; e.kayacan@uq.edu.au}}

\affil{University of Queensland, Brisbane, QLD 4067, Australia}

\begin{document}
 
\maketitle

\section*{Nomenclature}
{\renewcommand\arraystretch{1.0}
\noindent\begin{longtable*}{@{}l @{\quad=\quad} l@{}}
    $A$ & open-loop dynamics matrix \\
    $A_m$ & desired closed-loop dynamics matrix \\
    $\Delta A$ & model mismatch matrix \\
    $B_m$ & matched input gain matrix\\
    $B_{um}$ & unmatched input gain matrix \\
    $c$ & aerodynamic reference length, m \\
    $C$ & output matrix\\
    $C_{m_\alpha}$ & pitch stiffness coefficient, $\text{deg}^{-1}$ \\
    $C_{m_q}$ & pitch damping coefficient, $\text{(deg/s)}^{-1}$ \\
    $C_{m_{\delta_e}}$ & pitch control effectiveness coefficient, $\text{deg}^{-1}$ \\
    $C_{N_\alpha}$ & normal force gradient coefficient, $\text{deg}^{-1}$  \\
    $C_{N_{\delta_e}}$ & normal force due to elevator deflection coefficient,  $\text{deg}^{-1}$ \\
    $f_1(x(t),z(t),t))$ & matched uncertainty \\
    $f_2(x(t),z(t),t))$ & unmatched uncertainty \\
    $F$ & free-response matrix \\
    $G$ & forced response matrix \\
%    $H$ & hessian \\
    $I_2$ & y component of moment of inertia matrix, $\text{kg m}^2$ \\
    % $J$ & cost function \\
    $K$ & optimal gain matrix \\
    $m$ & number of control inputs \\
    $\bar{m}$ & mass, kg \\ 
	$n$ & number of states \\
% 	$n_p$ & number of prediction points \\
	$Q$ & output error weight \\
	$R$ & control weight \\
	$q$ & pitch rate, deg/s \\
	$\bar{q}$ & dynamic pressure, Pa \\
	$\bar{S}$ & aerodynamic reference area, $\text{m}^2$ \\
	$S$ & eigenvectors \\
	$u_{opt}$ & optimal control input \\
	$u_{ad}$ & adaptive control input \\
	$V$ & velocity, m/s \\
	$x$ & state vector \\
	$x_z$ & internal unmodelled dynamics \\
	$\tilde{x}$ & state prediction error \\
    $y_r$ & reference command \\
    $\alpha$ & angle of attack, deg \\
	$\lambda$ & eigenvalue \\
	$\Phi$ & state transition matrix \\
    $\omega_u$ & input gain \\
	$||\cdot||_2$ & euclidean norm \\ 
\end{longtable*}}

\section{Introduction}

Model predictive control (MPC) is an optimal control strategy where control input calculation is based on minimizing the predicted tracking error over a finite horizon that moves with time. This strategy has an advantage over conventional state feedback and output feedback controllers because it predicts the response of the system, rather than simply reacting to it. Therefore, MPC can offer improved performance in the presence of input and output constraints. Many implementations of MPC on aerospace vehicles appear in literature \cite{Eren2017}. Some of these include spacecraft and satellite attitude control \cite{chen2011design, wood2008model, hegrenaes2005spacecraft}, spacecraft rendezvous and docking \cite{weiss2015model}, helicopters \cite{ngo2015model}, and atmospheric re-entry \cite{van2006combined, pascucci2015model}.

There are many unique challenges associated with the control of hypersonic and supersonic re-entry vehicles, including 1) nonlinear and highly uncertain dynamics, 2) time-varying plant parameters, and 3) fast (and sometimes unstable) open-loop state dynamics, requiring a high control update rate. 
MPC was applied in simulation for the re-entry X-38 unpowered crew return vehicle (CRV), using quadratic programming at an update rate of 10 Hz \cite{van2006combined}. However, the CRV has stable open-loop longitudinal dynamics at high angles of attack \cite{horvath2004x}, allowing for the low update rate. In general, reusable boosters may not have stable re-entry open-loop dynamics \cite{eggers2003aerodynamic, chai2017fly}. Therefore, a higher control update rate might be required than what is computationally tractable with a conventional constrained quadratic program based MPC. 

A number of strategies exist to address the problem of computational tractability in MPC. Explicit MPC involves pre-computing some (or all) of the optimal control solutions and embedding them into multi-dimensional look-up tables for online access \cite{Lamburn2014, hegrenaes2005spacecraft}. However, this involves a tedious and labour-intensive process of compiling all the required solutions into tabular form. The complexity increases further if the model is time-varying or has uncertain parameters. 

A set of algorithms called efficient model predictive control  have emerged in recent literature, which reduce the number of prediction points to reduce the computational load \cite{Gibbens2011, Abdolhosseini2013, Lamburn2014}. One of these algorithms, called algebraic model predictive control (AMPC), involves using an eigendecomposition to calculate the state transition matrix, resulting in no truncation error regardless of the time step size \cite{Gibbens2011}. Consequently, an absolute minimum of one prediction point may be used without truncation error penalties in the prediction step. For the unconstrained case, quadratic programming is not required, and the optimal control solution for AMPC may be reduced into a closed-form expression analogous to a state feedback controller. 

One of the well-understood limitations of all MPC schemes is their reliance on an accurate model, which determines the quality of the predictions, and therefore, the optimal control input computed by the algorithm. The presence of unmodeled nonlinear dynamics, noisy data, as well as aerodynamic and gravimetric uncertainties all contribute to model inaccuracies. This note aims to provide a control scheme that alleviates this reliance on an accurate model by applying an adaptive control augmentation to the AMPC control law. Adaptive MPC has been investigated previously by incorporating a parameter estimation algorithm to identify the model used by the MPC \cite{fukushima2007adaptive,Adetola2009, kayacan, KayacanRSS}. These studies show that the control performance is initially conservative due to the large parametric uncertainty, but improves over time as the parameter estimates converge to their true values. However, persistent excitation is a requirement for convergence of the parameter estimator.

$\mathcal{L}_1$ adaptive control for matched uncertainties has also been combined with MPC and applied to the trajectory tracking problem of a quadrotor, resulting in improved experimental performance compared to non-predictive and non-adaptive approaches \cite{Pereida2018}. The MPC controller is used to compute the optimal reference command, which is then augmented by the $\mathcal{L}_1$ adaptive control law. In the presence of dynamic disturbances, the MPC-$\mathcal{L}_1$ controller was able to retain highly accurate tracking performance. Furthermore, the study showed that the $\mathcal{L}_1$-augmented non-predictive control schemes had more time delay compared to the MPC-$\mathcal{L}_1$ scheme.

Adaptive control methods that utilize an online parameter estimator have severe limitations when applied to hypersonic and supersonic re-entry. These limitations are due to the persistent excitation requirement as well as fast and time-varying system parameters. A study involving system identification methods for adaptive control of a hypersonic glider has shown that the convergence for the parameter estimation algorithm is too slow in the presence of the fast parametric variation in hypersonic flight \cite{creagh2011adaptive}. $\mathcal{L}_1$ adaptive control is the preferred augmentation for several reasons: 1) the transient performance can be known without requiring persistent excitation, 2) decoupling of robustness and adaptation rate allows for fast adaptation, and 3) a bounded time delay margin \cite{Banerjee2015}. For the reasons listed, $\mathcal{L}_1$ adaptive control has been implemented on aerospace systems such as quadrotors, aircraft, missiles, and hypersonic vehicles \cite{Pereida2019,Grodahl2011LAC,Cao2009,Banerjee2015,Banerjee2016}.

The main contributions of this note are: 1) Formulation of AMPC to control a linear time-varying (LTV) booster re-entry model, 2) presentation of an adaptive AMPC scheme, combining the AMPC optimal control law with an $\mathcal{L}_1$ augmentation for matched and unmatched uncertainties, 3) showing that the closed-form solution in unconstrained AMPC can be exploited to cast the state-space model into a partially closed-loop form amenable to adaptive control synthesis, and 4) showing that the AMPC-$\mathcal{L}_1$ scheme outperforms non-adaptive AMPC in off-design conditions. 

The note is as follows: Section \ref{sec:controller} presents the control methodologies for AMPC and AMPC-$\mathcal{L}_1$. In Section \ref{sec:results}, simulation results are discussed, and the control schemes are benchmarked against each other. A Monte Carlo analysis and robustness analysis are also conducted. Section \ref{sec:conclusions} details the final conclusions. The booster re-entry model is provided in the Appendix. 

%\newpage
%\clearpage

\section{Controller Design}
\label{sec:controller}

\subsection{Algebraic MPC}

% \paragraph{Remark 1:} Results from literature have revealed that the inclusion of terminal cost and constraints are crucial in ensuring the stability of constrained MPC \cite{Mayne2000}. However, for unconstrained AMPC, the optimal control law is a closed-form solution. Therefore, closed-loop stability may be assessed using pole-zero methods from classical control, negating the requirement of terminal cost and constraints in the formulation of Equation \eqref{eqn2.21}. 
AMPC is a model predictive control strategy that utilises an exact solution for the state transition matrix in its prediction step. The unconstrained, single prediction point form of AMPC is used in this study. The full derivation is not provided here, but may be found in \cite{Gibbens2011}. For the unconstrained case, the optimal control law can be obtained by minimizing the cost function with respect to the hypothesized control inputs. The result of the optimization yields the gain matrix acting on the predicted output errors. That optimal gain matrix is
\begin{equation}
	K = (G^{T}QG + R)^{-1}G^{T}Q,
	\label{eqn2.26}
\end{equation}
where $G$ is the forced response matrix, $Q$ is the weighting matrix penalising predicted output errors, and $R$ is the weighting matrix penalising control inputs. The matrix $K$ is the optimal set of gains acting on the error between the predicted response of the system and the reference command. 

The single prediction point AMPC makes the matrices the smallest possible, thereby allowing for the fastest update rates. For a single-input single-output (SISO) system, the predicted error and control activity weights become scalars. 
For a linear state-space system of the form
\begin{equation}
	\dot{x} = Ax + B_mu,
	\label{eqn:state_space_linear}
\end{equation}
the free and forced response matrices are
\begin{align}
    \label{eq:F}
    F &= C\Phi(\delta t)  \\
    G &= C A^{-1} \Big(\Phi(\delta t) - I \Big)B_m,
    \label{eq:G}
\end{align}
where $\delta t$ is the prediction time horizon, $A$ is the state dynamics matrix, $B_m$ is the control input matrix, $C$ is the output matrix, $F$ is the free-response matrix with $F \in \mathbb{R}^{1 \times n}$, and $G \in \mathbb{R}$. The distinctive feature of AMPC is that the state transition matrix $\Phi(\delta t) $ is evaluated by computing the exact matrix exponential.
The exact matrix exponential is computed using the eigenvalues and eigenvectors of the state dynamics matrix instead of a Taylor series expansion \cite{Gibbens2011}. For a given time interval, the state transition matrix can be expressed as
\begin{equation}
	\Phi(\delta t) = e^{A(\delta t)},
 \label{eqn3.2}
\end{equation}
where
\begin{equation}
	e^{A(\delta t)} =  S \; \text{diag}(e^{\lambda_{1}(\delta t)},\dots, e^{\lambda_{k}(\delta t)}) \;%\left(
	%\begin{array}{cccc}
	%e^{\lambda_{1}(\delta t_{i+1} - \delta t_{i})}&0&\cdots&0\\
	%0&e^{\lambda_{2}(\delta t_{i+1} - \delta t_{i})}& & \\
	%\vdots& & \ddots & \vdots\\
	%0 &0 & \cdots & e^{\lambda_{n_p}(\delta t_{i+1} - \delta t_{i})}\\
	%\end{array}
	%\right)
	S^{-1},
 \label{eqn3.3}
\end{equation}
with $S$ defining a set of eigenvectors corresponding to the system eigenvalues $\lambda_{1}$ to $\lambda_{k}$. There is no truncation error in this formulation of the matrix exponential. No approximations are made, as a direct solution of the state transition matrix can be found. Any resulting inaccuracies that may stem from the state transition matrix will therefore only be subject to the precision of the model \cite{Gibbens2011}.

The optimal control law is formulated using the optimal gain $K$
\begin{equation}
    \label{eq:siso_mpc_control}
    u_{opt} = K (y_r - Fx).
\end{equation}
This formulation is used in both the baseline AMPC and AMPC-$\mathcal{L}_1$ controllers for the purposes of this study. 

 Equations (\ref{eqn2.26}) to (\ref{eq:siso_mpc_control}) are sufficient if the system under control is LTI. For time-varying systems, $K(t)$, $F(t)$, and $G(t)$ must be re-computed at every control update based on the latest system parameters. That is,
        \begin{align}
            F(t) &= C\Phi(\delta t, t),  \\
            G(t) &= C A(t)^{-1} \Big(\Phi(\delta t, t) - I \Big)B_m(t), \\
            K(t) &= \Big(G(t)^{T}QG(t) + R \Big)^{-1}G(t)^{T}Q. \label{eq:K(t)}
        \end{align}
        
This produces errors in the AMPC prediction, as the free and forced response matrices are assumed to be constant throughout the prediction horizon, when they are time-varying in reality. 
These modelling errors result in errors in $A$ and $B_m$. However, the rate of variation of the parameters is slow relative to the prediction time horizon. Therefore, the amount of uncertainty throughout the prediction horizon is very small. Furthermore, the $\mathcal{L}_1$ adaptive augmentation, which is introduced in the subsequent section, is able to compensate for matched and unmatched uncertainties. The control performance under these assumptions is evaluated in Section \ref{sec:results}.

\subsection{Algebraic MPC with $\mathcal{L}_1$ adaptive augmentation}
An $\mathcal{L}_1$ adaptive controller with piecewise-constant adaptive law for matched and unmatched nonlinear uncertainties is used as the adaptive augmentation for AMPC \cite{Xargay2014, Banerjee2015}. The piecewise-constant adaptive law avoids the risk of numerical stiffness associated with the selection of high adaptive gains in MRAC-type adaptive control schemes.

 The AMPC is used to compute the desired closed-loop dynamics $A_m$ and the optimal gain $K$, which are both used in the $\mathcal{L}_1$ adaptive augmentation. Extensive proofs relating to $\mathcal{L}_1$ adaptive architectures are provided in \cite{Xargay2014} and \cite{Hovakimyan2011}; therefore, only the main results along with a derivation of the partial closed-loop system using AMPC results will be presented in this section.
 
 The architecture of $\mathcal{L}_1$ adaptive controllers consists of a state predictor, an adaptive law, and a control law, as illustrated in Figure \ref{fig:adaptive_control_blocks}. The goal of adaptive control is to modify the control law based on the uncertainties in the plant. The state predictor uses the estimated uncertainties and control inputs to predict a future state. The error between the predicted state and the actual state is used in the next adaptive step to drive the adaptive law, which modifies the estimated uncertainties. The control law is then modified based on the estimated uncertainties.

\tikzstyle{block} = [draw, rectangle, 
    minimum height=3em, minimum width=7em, line width = 0.3mm]
\tikzstyle{sum} = [draw, circle, node distance=2cm]
\tikzstyle{input} = [coordinate]
\tikzstyle{output} = [coordinate]
\tikzstyle{pinstyle} = [pin edge={to-,thin,black}]
\usetikzlibrary{fit}
\usetikzlibrary{arrows.meta}
\tikzset{
  FARROW/.style={line width=0.4mm, arrows={-Latex[angle=40:2.8mm]}}
}

% The block diagram code is probably more verbose than necessary
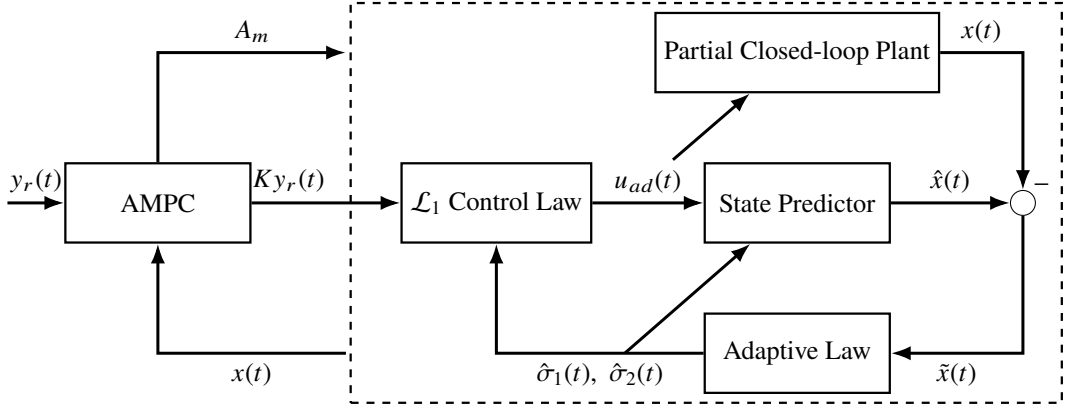
\begin{figure}
\centering
\begin{tikzpicture}[auto, node distance=2cm,>=latex']
    % We start by placing the blocks
    \node [input, name=input] {};
    %\node [sum, right of=input] (sum) {};
    \node [block, right of=input] (ampc) {AMPC};
    \node [input, right of=ampc, node distance=2.5cm] (in2) {};
    \node [block, right of=in2] (controller) {$\mathcal{L}_1$ Control Law};
    \node [output, below of=in2] (feedback) {};
    \node [input, above of=in2] (in3) {};
    \node [output, right of=in2, node distance=0.18cm] (in5) {};
    \node [block, right of=controller, node distance=4cm] (predictor) {State Predictor};
    % We draw an edge between the controller and system block to 
    % calculate the coordinate u. We need it to place the measurement block. 
    %\draw [->] (controller) -- node[name=u] {$u$} (plant);
    \node [block, above of=predictor] (plant) {Partial Closed-loop Plant};
    \node [sum, right of=predictor, node distance=3cm] (sum_o) {};
    \node [output, right of=sum_o, node distance=0.4cm] (out2) {};

    \node [block, below of=predictor] (adaptive_law) {Adaptive Law};
    \node [input, right of=plant] (out) {};
    \node[draw,dashed,line width=0.3mm,fit=(controller) (predictor) (plant) (sum_o)(in5) (adaptive_law) (out2)] {};

    % Once the nodes are placed, connecting them is easy. 
    \draw [FARROW] (input) -- node {$y_r(t)$} (ampc);
    \draw [FARROW] (ampc) -- node [near start] {$K y_r(t)$} (controller);
    \draw [FARROW] (ampc) |- node [near end] {$A_m$} (in3);
    %\draw [->] (controller) |- node {} (plant);
    \draw [FARROW] (controller) -- node [name=u] {$u_{ad}(t)$}(predictor);
    \draw [FARROW] (u) -- node {} (plant);
    \draw [FARROW] (predictor) -- node [name=x] {$\hat{x}(t)$}(sum_o);
    \draw [FARROW] (plant) -| node[pos=0.99] {$-$} 
        node [near start] {$x(t)$} (sum_o);
    \draw [FARROW] (sum_o) |- node [near end] {$\tilde{x}(t)$} (adaptive_law);
    \draw [FARROW] (adaptive_law) -| node [near start, name=sigma] {$\hat{\sigma}_1(t), \; \hat{\sigma}_2(t)$} (controller);
    \draw [FARROW] (sigma) -- node {} (predictor);
    \draw [FARROW] (feedback) -| node [near start] {$x(t)$} (ampc);
\end{tikzpicture}
\caption{\label{fig:adaptive_control_blocks} AMPC-$\mathcal{L}_1$ architecture for matched and unmatched uncertainties $\hat{\sigma}_1(t), \hat{\sigma}_2(t)$. The dashed box encloses the $\mathcal{L}_1$ adaptive augmentation, which interfaces with AMPC via the desired closed-loop dynamics matrix $A_m$, reference command $K y_r$, and state $x(t)$. }
\end{figure}

\subsubsection{Derivation of Partial Closed-loop System}
The full control input comprises of the optimal control input computed by AMPC and adaptive component as follows:
\begin{equation}
    \label{eq:full_control}
    u(t) = u_{opt}(t) + u_{ad}(t),
\end{equation}
where $u_{opt}$ is the optimal control input determined by AMPC and $u_{ad}$ is the adaptive component. Combining Equations \eqref{eqn:state_space_linear},  \eqref{eq:siso_mpc_control}, and \eqref{eq:full_control} yields 
\begin{equation}
    \dot{x}(t) = Ax(t) + B_m \bigg(K \Big(y_r(t) - F x(t) \Big) + u_{ad}(t) \bigg)
\end{equation}
which may be placed into the partial closed-loop form
\begin{equation}
    \label{eq:partial_closed}
    \dot{x}(t) = (A - B_m K F)x(t) + B_m \Big(K y_r(t) + u_{ad}(t) \Big).
\end{equation}
The optimal AMPC gain matrix $K$ and predicted free-response $F$ is used to make the partial closed-loop system Hurwitz, yielding the desired closed-loop dynamics matrix $A_m = A - B_m K F$.  Furthermore, moving $K y_r$ into the adaptive component yields
\begin{equation}
    \label{eq:partial_full}
    \dot{x}(t) = A_m x(t) + B_m u_{ad}(t),
\end{equation}
where $u_{ad}$ is the adaptive control component that includes $K y_r$. In the presence of unmodelled nonlinear dynamics, matched and unmatched uncertainties as well as uncertain input gain, Equation \eqref{eq:partial_full} becomes
\begin{align}
\dot{x}(t) &= A_m x(t) + B_m \bigg(\omega_u u_{ad}(t) + f_1 \Big(x(t),z(t),t \Big) \bigg) + B_{um} f_2 \Big(x(t),z(t),t \Big), \;\;\;\; x(0) = x_0 \\
z(t) &= g_o \Big(t,x_z(t) \Big) \\
\dot{x}_z(t) &= g \Big(x_z(t),x(t),t \Big), \;\;\;\; x_z(0) = x_{z0} \\
y(t) &= C x(t)
\end{align}
where $B_{um}$ and $B_{m}$ are the unmatched and matched input gain, respectively. The unmatched input gain, $B_{um} \in \mathbb{R}^{n\times(n-m)}$, is a matrix such that $B_m^T B_{um} = 0 $ and $\text{rank}([B_m B_{um}]) = n $. The terms $f_1$ and $f_2$ are the matched and unmatched uncertainties, respectively, while $x_z$ is the internal unmodelled dynamics. $z$ is the output of the unmodelled dynamics. This adaptive control method operates under a number of assumptions, detailed below \cite{Hovakimyan2011}. Let $X \triangleq [x^T, z^T]^T$ and $f_{i}(t,X) \triangleq f_{i}(t, x, z)$, $i = 1,2$.

\paragraph{Assumption 1: Boundedness of the matched (\bm{$f_{1}(t,0,0)$}) and unmatched ($\bm{f_{2}(t,0,0)}$) uncertainties.} There exists $B_i > 0$, such that $||f_i(t,0)||_\infty \leq B_i$ holds for all $t \geq 0$ and $i = 1,2$.
 
\paragraph{Assumption 2: Semiglobal Lipschitz condition on $\bm{f_{1}}$ and $\bm{f_{2}}$.} For $i=1,2$ and arbitrary $\delta > 0$, there exist positive $K_{1_\delta}$, $K_{2_\delta}$, such that
\[ || f_i(t, X_1) - f_i(t,X_2)  ||_\infty \leq K_{i_\delta} || X_1 - X_2  ||_\infty, \; \;  i=1,2,  \]
for all $|| X_j ||_\infty \leq \delta, j=1,2,$ uniformly in $t$.
\paragraph{Assumption 3: BIBO stability of unmodelled dynamics.} The $x_z$-dynamics are BIBO stable with respect to both initial conditions $x_{z0}$ and input $x(t)$, i.e., there exist $L_z$, $B_z > 0$ such that for all $t \geq 0 $ 
\[ ||z_t||_{\mathcal{L}_\infty} \leq L_z || x_t ||_{\mathcal{L}_\infty} + B_z, \]
 where $z_t$ and $x_t$ are $z(t)$ and $x(t)$ evaluated at time $t$.
\paragraph{Assumption 4: Partial knowledge of system input gain.} The system input gain matrix $\omega$ is assumed to be an unknown (nonsingular) strictly row-diagonally dominant matrix with the sign of $\omega_{ii}$ being known. Also, assume that there exists a known compact convex set $\Omega$, such that $\omega \in \Omega \subset \mathbb{R}^{m \times m}$, and that a nominal system input gain $\omega_0 \in \Omega$ is known.

%subsubsection{$\mathcal{L}_1$-norm Sufficient Condition for Stability }
%A central requirement in the design of $\mathcal{L}_1$ adaptive controllers is the $\mathcal{L}_1$-norm sufficient condition for stability. The matched and unmatched transmission have transfer functions defined as
%\begin{align*}
% H_{xm}(s) &\triangleq (sI_n - A_m)^{-1}B_m,  \\
% H_{xum}(s) &\triangleq (sI_n - A_m)^{-1}B_{um}, \\  
% H_m(s) &\triangleq CH_{xm}(s), \text{and}\\
% H_{um}(s) &\triangleq CH_{xum}(s). 
%\end{align*}
%Including a low pass filter $C(s)$ to the adaptive control input results in the transfer functions
%\begin{align*}
%    G_m(s) &\triangleq H_{xm}(s)(I_m - C(s)), \text{and} \\
%    G_{um}(s) &\triangleq (I_n - H_{xm}(s) C(s)H_m^{-1}(s)C)H_{xum}(s)
%\end{align*}
%\begin{align*}
% x_{in}(s) &\triangleq (sI_n - A_m)^{-1} x_0, \\
% \rho_{in} &\triangleq ||s(sI_n - A_m)^{-1}||_{\mathcal{L}_1} \rho_0,\\
% L_{i\delta} &\triangleq \frac{\bar{\delta}(\delta)}{\delta} K_{i \bar{\delta}(\delta)},\\
% \bar{\delta}(\delta) &\triangleq \max \{ \delta + \bar{\gamma}, L_z(\delta + \bar{\gamma}_1) + B_z \}, \\
% C(s) &\triangleq \frac{\omega}{s + \omega}, \\
%\end{align*}
%hold, where $\bar{\gamma}$ is an arbitrarily small positive constant. The $\mathcal{L}_1$-norm condition is defined as
%%%\begin{equation}
%||G_m(s)||_{\mathcal{L}_1} + ||G_{um}||_{\mathcal{L}_1} \el_0 < \frac{\rho_r - ||H_{xm}(s)C(s)K_g(s)||_{\mathcal{L}_1} ||r||_{\mathcal{L}_\infty} - \rho_{in}}{L_1{\rho_r}\rho_r + B_0}
%\end{equation}

\subsubsection{State Predictor}
The state predictor is used to predict the system response based on the current estimates of uncertainties $\hat{\sigma_1}$ and $\hat{\sigma_2}$. The error in prediction drives the adaptive law, which updates the uncertainty estimates. The state predictor has the form

\begin{align}
    \dot{\hat{x}}(t) &= A_m \hat{x}(t) + B_m \Big(u_{ad}(t) + \hat{\sigma}_1(t) \Big) + B_{um}\hat{\sigma}_2(t), \;\;\;\; \hat{x}(0) = x_0 
    \label{eq:state_predictor}
\end{align}
where $\hat{\sigma}_1(t)$ and $\hat{\sigma}_2(t)$ are the current adaptive estimates for the matched and unmatched uncertainties ($f_1(x(t),z(t),t))$ and $f_2(x(t),z(t),t))$), respectively. For the purposes of this note, the state predictor is implemented in simulation to update the predictions at the same rate as the adaptive law.

\subsubsection{Adaptive Law}
A piecewise-constant adaptive law, first introduced in \cite{Cao2009}, is used to update the estimated uncertainties $\hat{\sigma}_1(t)$ and $\hat{\sigma}_2(t)$. The piecewise-constant update law does not require tuning of an adaptive gain, but rather utilizes the adaptation sampling time $T_s$, which may be set arbitrarily small subject only to CPU restrictions. For the $i$th adaptation, the adaptative law is

\begin{equation}
    \hat{\sigma}(iT_s) = -B^{-1}\Phi_{ad}^{-1}(T_s)\mu(iT_s),\;\;\;\; t \in [iT_s,(i+1)T_s],
    \label{eq:adaptive_law}
\end{equation}
where
\begin{align}
    \hat{\sigma}(iT_s) &= 
    \begin{bmatrix}
    \hat{\sigma}_1(iT_s) \\ 
    \hat{\sigma}_2(iT_s) \label{eq:adaptive_law1}
    \end{bmatrix} \\
    \Phi_{ad}(T_s) &\triangleq A_m^{-1} ( e^{A_m T_s} - \mathbb{I}_n), \label{eq:adaptive_law2}\\
    \mu(iT_s) &= e^{A_m T_s} \tilde{x}(iT_s), \label{eq:adaptive_law3}\\
    \tilde{x}(t) &= \hat{x}(t) - x(t), \; \text{and} \label{eq:adaptive_law4}\\
    B &= [B_m\; B_{um}] \label{eq:adaptive_law5}.
\end{align}
For an LTI system, Equations \eqref{eq:adaptive_law2} and \eqref{eq:adaptive_law5} would need only to be computed once as the matrices $A_m$ and $B$ are time invariant. In the LTV case, Equations \eqref{eq:adaptive_law2} to \eqref{eq:adaptive_law5} must be re-computed at every adaptive step as part of the successive linearization process.  

\subsubsection{Control Law}

The matched and unmatched transmission transfer functions are defined in the Laplace domain as
\begin{align*}
 H_m(s) &\triangleq C H_{xm}(s), \\
 H_{um}(s) &\triangleq C H_{xum}(s),
\end{align*}
where
\begin{align*}
H_{xm}(s) &\triangleq (s\mathbb{I}_n - A_m)^{-1}B_m,\; \text{and}\\
H_{xum}(s) &\triangleq (s\mathbb{I}_n - A_m)^{-1}B_{um}.
\end{align*}

These transfer functions are used to formulate the adaptive control law through compensation of the estimated unmatched uncertainty. The adaptive control law in the Laplace domain is
\begin{equation}
    u_{ad}(s) = C(s) \Big(\hat{\sigma}_1(s) + H_m^{-1}(s) H_{um}(s) \hat{\sigma}_2(s) - K y_r(s) \Big)
    \label{eq:control_law}
\end{equation}
where $C(s)$ is a user-specified low-pass filter and $K$ is the AMPC optimal gain matrix used to track references. The inverse of the matched transmission, $H_m^{-1}(s)$, is unstable for non-minimum phase systems. The angle of the attack dynamics of a fly back booster during re-entry are non-minimum phase. Therefore, the inverse DC gain of $H_m(s)$ is used instead to calculate $H_m^{-1}(s) H_{um}(s) = -(CA_m^{-1}B_m)^{-1}H_{um}(s)$, a method first proposed in \cite{che2012}. A property of transfer functions is that the inverse DC gain of $H_m(s)$ is the steady-state value of $H_m^{-1}(s)$. In the numerical simulation, $H_{um}(s)$ is implemented as a state-space model.

\subsubsection{Low-pass Filter and $\mathcal{L}_1$-norm sufficient condition}
In $\mathcal{L}_1$ adaptive control architectures, the low-pass filter is used to decouple the robustness of an adaptive system from the adaptation rate \cite{Hovakimyan2011}. Therefore, arbitrarily fast adaptations may be specified, subject only to CPU restrictions, while the low-pass filter is designed to bound the stability margins. In this study, the low-pass filter $C(s)$ is a first-order system of the form

\begin{equation}
    C(s) = \frac{\omega_c}{s + \omega_c},
\end{equation}
with cutoff frequency chosen based on the trade-off between performance and robustness. 
                
        Additionally, the cutoff frequency must be chosen to satisfy the $\mathcal{L}_1$-norm condition for closed-loop stability, which is stated as follows \cite{Hovakimyan2011}.
        
        Let $\rho_{in} \triangleq ||s(s \mathbb{I}_n - A_m)^{-1}||_{\mathcal{L}_1} \rho_0$ and let $x_{in}(t)$ be the inverse Laplace transform of $x_{in}(s) \triangleq (s\mathbb{I}_n - A_m)^{-1}$. Since $A_m$ is Hurwitz due to AMPC and $x_0$ is bounded, then $||x_{in}||_{\mathcal{L}_1} \leq \rho_{in}$. Furthermore, define
        
        \begin{align}
        L_{i\delta} & \triangleq \frac{\bar{\delta}(\delta)}{\delta} K_{i\bar{\delta}(\delta)}, \\
        \bar{\delta}(\delta) & \triangleq \text{max}\{\delta + \bar{\gamma_1}, L_z(\delta+\bar{\gamma_1}) + B_z\},
        \end{align}
        where $K_{i\delta}$ is defined in Assumption 2, $\bar{\gamma}_1$ is an arbitrarily small positive constant, and $L_z, B_z$ are defined in Assumption 3. For a given $\rho_0$, there exists $\rho_r > \rho_{in}$ such that the following inequality holds:
        
        \begin{equation}
            ||G_m(s)||_{\mathcal{L}_1} + ||G_{um}(s)||_{\mathcal{L}_1} \ell_0 < \frac{\rho_r - ||H_{xm}(s) C(s) K_g(s)||_{\mathcal{L}_1} ||r||_{\mathcal{L}_\infty} - \rho_{\text{in}}}{L_{1\rho_r}\rho_r + B_0},
        \end{equation}
        where
        \begin{align} 
        G_m(s) &\triangleq H_{xm}(s) \Big(\mathbb{I}_m - C(s) \Big), \\
        G_{um}(s) & \triangleq \Big(\mathbb{I}_n - H_{xm}(s)C(s) H_m^{-1}(s)C \Big)H_{xum}(s), 
        \end{align}
        while
        \[
            \ell_0 \triangleq \frac{L_{2\rho_r}}{L_{1\rho_r}},  \; \;  b_0 \triangleq \max \bigg\{B_{10},\frac{B_{20}}{\ell_0}\bigg\}.
        \]
        In the AMPC-$\mathcal{L}_1$ scheme, $K_g(s) = K$ based on the optimal gain calculated using Equation (\ref{eqn2.26}).

\section{Simulation, Results, and Analysis}
\label{sec:results}
\subsection{Simulation Setup}
This section presents a performance and robustness comparison between AMPC-$\mathcal{L}_1$ and baseline, non-adaptive AMPC. The structure of the plant model used for numerical simulation is
\begin{align}
    \label{eq:sim_model}
    \dot{x}(t) &= \Big(A(t) + \Delta A(t) \Big) x(t) + B_m(t) \omega_u \Big(u_{ad}(t) - K(t) F(t) x(t) \Big) + d(t),
\end{align}
where $\Delta A(t)$ is the plant model mismatch, $d(t)$ is the disturbance vector, and $\omega_u$ is the input gain. The plant model mismatch is used to represent uncertainties in aerodynamic parameters while the input gain may be reduced to represent loss of control authority in the actuator loop. For baseline AMPC, $u_{ad}(t) = K(t)y_r(t)$.

Several test cases are developed to benchmark the AMPC-$\mathcal{L}_1$ against the baseline AMPC in this note. These cases include under nominal conditions, reduced input gain, severe model mismatch, and disturbances as shown in Table \ref{simulation_cases}. The nominal condition in case 1 contains no uncertainties and assumes perfect knowledge of the plant. This is to validate that the AMPC-$\mathcal{L}_1$ behaves exactly like non-adaptive AMPC when there are no uncertainties. A Monte Carlo analysis is conducted to obtain a complete view of the controllers' performance properties under uncertain conditions.

\begin{table}[h]
	\centering
	\small
	\caption{\label{simulation_cases} Simulation cases for benchmarking AMPC-$\mathcal{L}_1$}
	\begin{tabular}{lllllll}
		\hline
		Simulation &  Description \\ \hline
		Case 1 	& Nominal case  \\ 
		Case 2 &  Reduced input gain ($40\% \; \omega_u$)	\\
		Case 3 &  Reduced input gain ($30\% \; \omega_u$ and model mismatch) \\ 
		Case 4 &  Disturbances \\ 
		Case 5 &  Monte Carlo simulation \\\hline
	\end{tabular}
\end{table} 

The AMPC design parameters used for this study are shown in Table \ref{design_params_mpc}, while design parameters for $\mathcal{L}_1$ adaptive control is shown in Table \ref{design_params}. The simulation is initialized with $\alpha = 2$ to represent a post stage separation maneuver, and the control update rate is set to 200 Hz. To compare the control performance of AMPC and AMPC-$\mathcal{L}_1$ quantitatively, the tracking error norm is used as a performance metric, defined using

\begin{equation}
    ||e||_2 = \sqrt{ \int | y_r(t) - x(t)|^2 \text{dt}}.
\end{equation}

\begin{table}[h]
	\centering
	\small
	\caption{\label{design_params_mpc} Design parameters for AMPC}
	\begin{tabular}{lllllll}
		\hline
		Parameter & Description & Value \\ \hline
		Q 	& Tracking error penalty weight &  0.99 \\ 
		R &  Control activity penalty weight & 	0.001\\
		$\delta t_{p}$ & Prediction horizon &  0.5 seconds\\ \hline
		
	\end{tabular}
\end{table} 

\begin{table}[h]
	\centering
	\small
	\caption{\label{design_params} Design parameters for $\mathcal{L}_1$ adaptive control}
	\begin{tabular}{lllllll}
		\hline
		Parameter & Description & Value \\ \hline
		$\omega_c$ 	& Low-pass filter cutoff frequency &  20 Hz \\ 
		$\delta t_{ad}$ &  Adaptive update time step & 	0.005 seconds \\ \hline
	\end{tabular}
\end{table} 
\subsection{Discussion of Results}
\begin{figure}[h!]
	\centering
	\includegraphics[width=\textwidth]{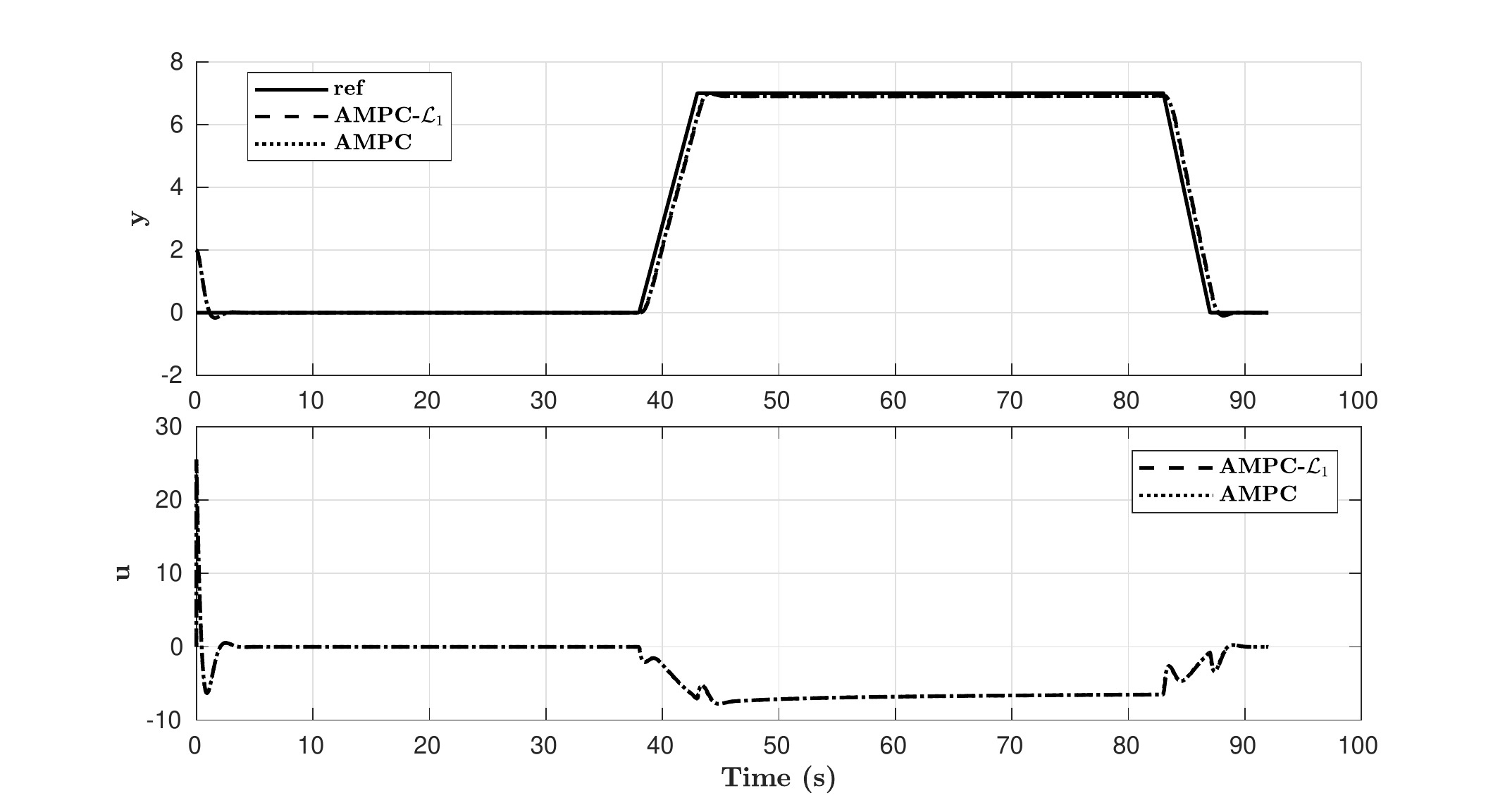}
	\caption{\label{fig:S1_plot} Tracking performance and control activity for case 1. $||e||_2 = 83$ for baseline AMPC and $||e||_2 = 88$ for AMPC-$\mathcal{L}_1$. The results are nearly identical, resulting in what appears to be a single dotted, dashed line.}
\end{figure}

\paragraph{Case 1 - Nominal Conditions:}
\label{sec:nominal}
The results for the nominal case are presented in Figure \ref{fig:S1_plot}. Under nominal conditions, the controllers generate near-identical control signals, resulting in very similar tracking-error norms of $||e||_2 = 83$ for baseline AMPC and $||e||_2 = 88$ for AMPC-$\mathcal{L}_1$. 

Firstly, this implies that the modelling errors from assuming that the free and forced response matrices are constant throughout the prediction horizon do not produce significant performance degradation for this system model.

Secondly, this shows that the $\mathcal{L}_1$ adaptive augmentation effectively lies dormant when there are negligible uncertainties in the system modelling or input gain. The fly back booster re-entry $\alpha$ commands are accurately tracked by both controllers, showing the viability of AMPC for the longitudinal control of re-entry vehicles.

\begin{figure}[h!]
	\centering
	\includegraphics[width=\textwidth]{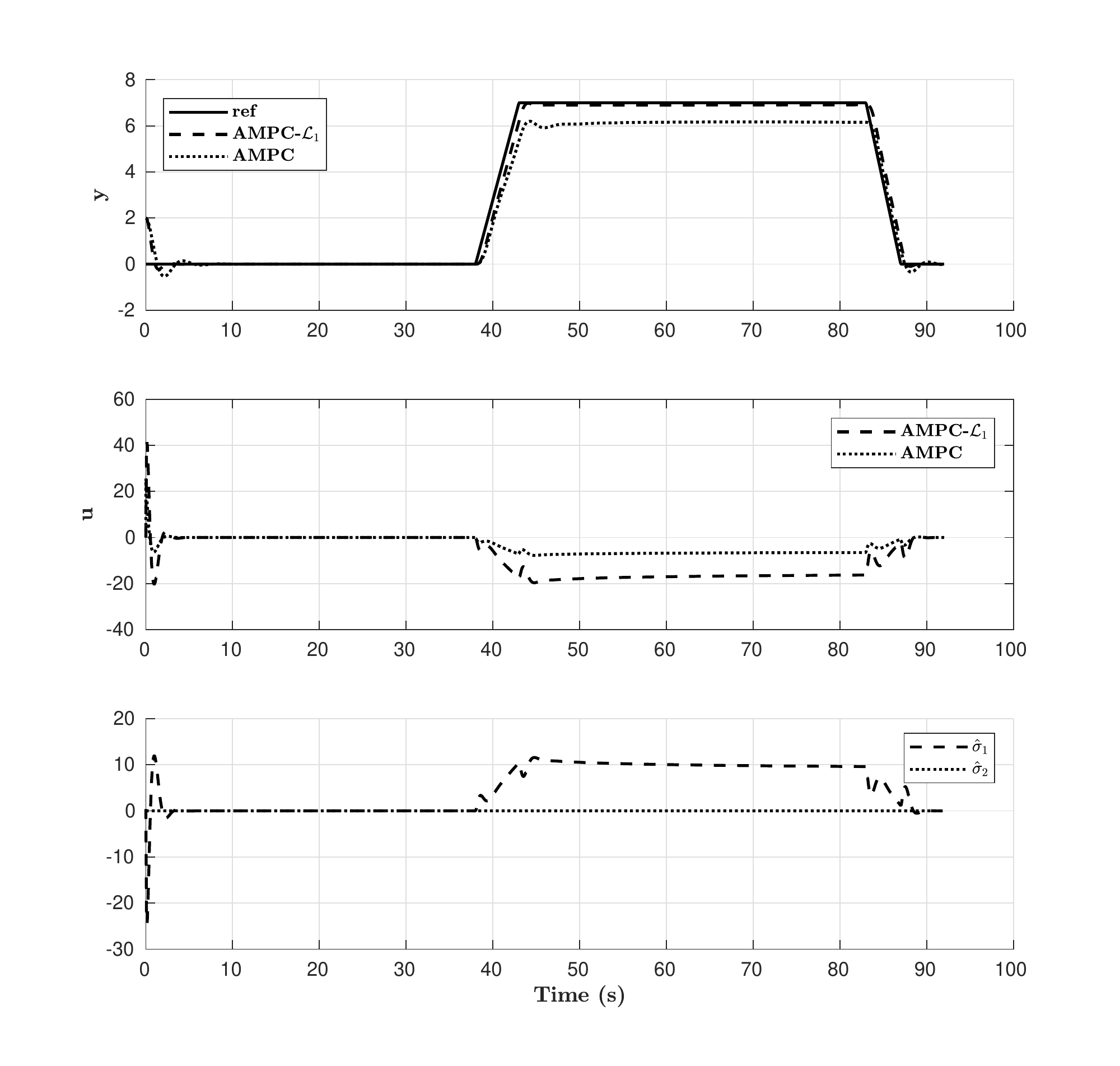}
	\caption{\label{fig:S2_plot} Tracking performance and control activity for case 2. $||e||_2 = 202$ for baseline AMPC and $||e||_2 = 90$ for AMPC-$\mathcal{L}_1$. The estimated uncertainties are dominated by the matched uncertainties, compensating for the uncertain input gain. }
\end{figure}

\paragraph{Case 2 - 40\% Reduced Input Gain:}
The tracking performance and control activity of both controllers under reduced input gain are shown in Figure \ref{fig:S2_plot}. The steady-state tracking performance of baseline AMPC is degraded, showing a steady-state error of one degree less than the reference command. This is because the AMPC algorithm computes the optimal control without any knowledge of the reduced input gain. Therefore, the AMPC bases its optimization on inaccurate predictions, yielding a control input that would reach the reference set point under nominal conditions, but results in steady-state error under reduced input gain. The transient performance of AMPC is also slightly degraded, with a larger overshoot and settling time compared to the nominal case.

On the other hand, the AMPC-$\mathcal{L}_1$ augmented controller retains the good performance from case 1 by adequately compensating for the reduced input gain. This is done by estimating the matched and unmatched uncertainties according to the piece-wise constant adaptive law and compensating for them by requesting more active control according to the control law. The tracking error norm for baseline AMPC is over twice compared to that of AMPC-$\mathcal{L}_1$.

\begin{figure}[h!]
	\centering
	\includegraphics[width=\textwidth]{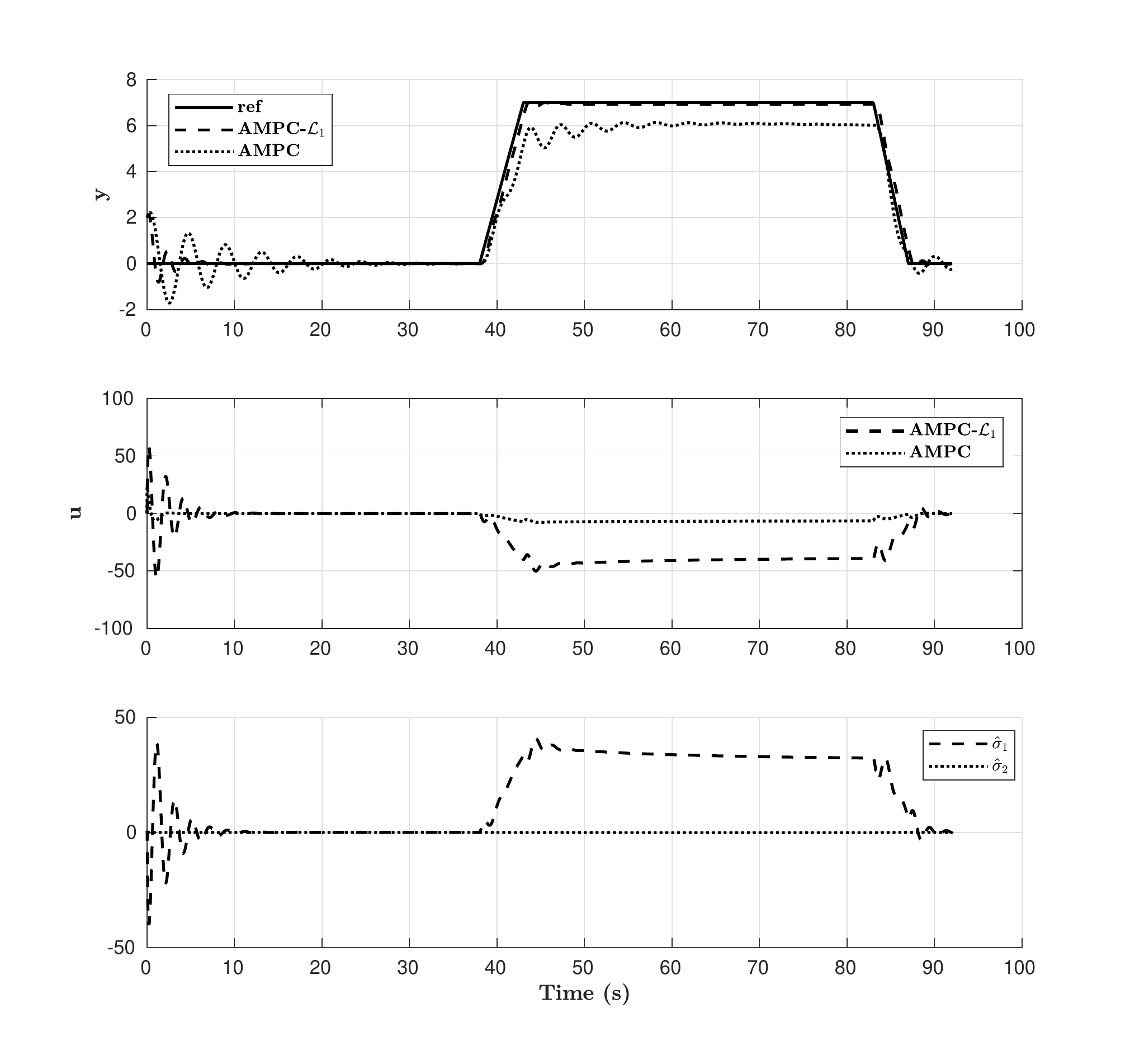}
	\caption{\label{fig:S4_plot} Tracking performance and control activity for case 3. $||e||_2 = 258$ for baseline AMPC and $||e||_2 = 81$ for AMPC-$\mathcal{L}_1$. The estimated matched uncertainties dominate the controller for this case. }
\end{figure}

\newpage
\clearpage
\paragraph{Case 3 - 30\% Reduced Input Gain and Model Mismatch:}
To assess the controllers' performance and robustness under model uncertainties, a severe, time-dependent model mismatch is imposed in the system. This includes reduced pitch rate damping $M_q$, reduced static stability $M_\alpha$, increased pitch rate to the angle of attack rate mapping, and increased normal force dependence on $\alpha$. This results in the model mismatch matrix
\begin{equation}
    \label{eq:model_mismatch}
    \Delta A(t) = 
    \begin{bmatrix}
    -0.8 M_q (t) & 0.8 M_\alpha (t) \\
    0.3 & 0.7
    \end{bmatrix}.
\end{equation}
The performance of both controllers may be seen in Figure \ref{fig:S4_plot}. There is severe degradation in the performance of the baseline AMPC, marked by oscillatory behaviour in the transient response and a steady-state error of one degree. In contrast, the AMPC-$\mathcal{L}_1$ can suppress the oscillations and retain very small steady-state error. The control activity of AMPC-$\mathcal{L}_1$ can be seen to be actively compensating for the model mismatch and reduced input gain. Quantitatively, the tracking error norm $||e||_2$ of the baseline AMPC controller is over three times higher than the AMPC-$\mathcal{L}_1$ controller.

\begin{figure}[h!]
	\centering
	\includegraphics[width=\textwidth]{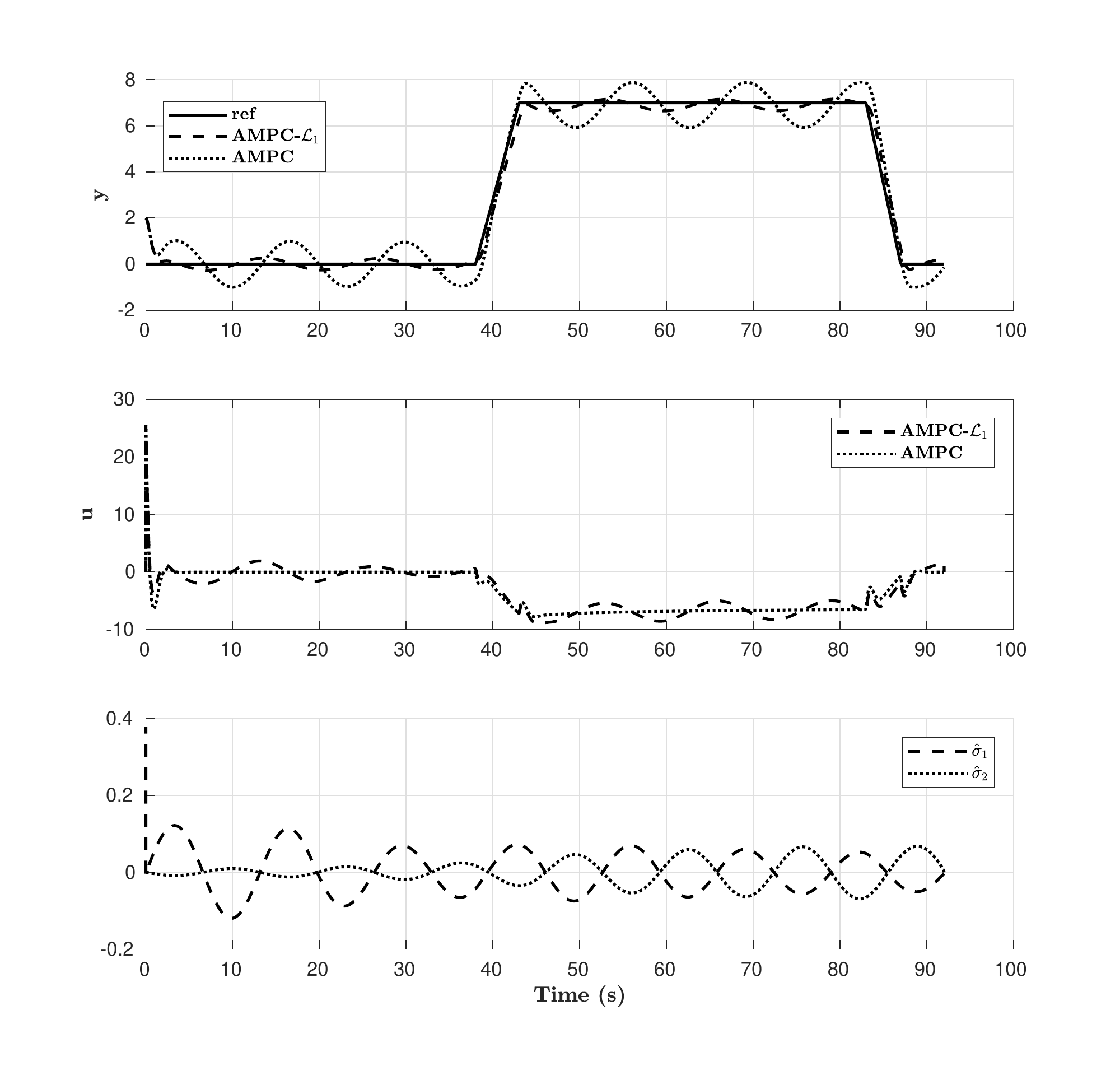}
	\caption{\label{fig:S5_plot} Tracking performance and control activity for case 4. $||e||_2 = 223$ for baseline AMPC and $||e||_2 = 89$ for AMPC-$\mathcal{L}_1$. The estimated matched and unmatched uncertainties are equally dominant in the case of disturbance rejection.}
\end{figure}

 \newpage
 \clearpage
\paragraph{Case 4 - Disturbances:}
To assess the robustness of both controllers to disturbances, a sinusoidal disturbance is imposed on the system defined in Equation \eqref{eq:sim_model}, defined by
\begin{equation}
    d(t) = 
    \begin{bmatrix}
    0 \\ 2\sin(1.5t/ \pi)
    \end{bmatrix}.
\end{equation}
The results from imposing the sinusoidal disturbance are shown in Figure \ref{fig:S5_plot}. These figures show that the AMPC-$\mathcal{L}_1$ has a lower tracking error norm than baseline AMPC in the presence of disturbances, evident by the AMPC-$\mathcal{L}_1$'s superior suppression of oscillations.
%\begin{figure}[h!]
	%\centering
	%\includegraphics[width=\textwidth]{S5_u.eps}
	%\caption{\label{fig:S5_u} Control activity of case 4}
%\end{figure}

\paragraph{Case 5 - Monte Carlo Analysis:}
\label{sec:monte_carlo}
Although cases 1 to 4 clearly show that the AMPC-$\mathcal{L}_1$ can provide better control performance when compared to baseline AMPC under some off-design conditions, a Monte Carlo analysis is conducted to establish a more systematic assessment of the controllers' performance under uncertain conditions. The uncertain parameters are assumed to be normally distributed, with standard deviations shown in Table \ref{uncertain_params}.

\begin{table}[h!]
	\centering
	\small
	\caption{\label{uncertain_params} Uncertain parameters for Monte Carlo simulation}
	\begin{tabular}{lllllll}
		\hline
		Uncertain Parameters & Description & Standard Deviation \\ \hline
		$\delta M_q$ 	& Pitch damping &  $\sigma(M_q) = 0.25 M_q$ \\ 
		$\delta M_\alpha$ &  Pitch stiffness & 	$\sigma(M_\alpha) = 0.25 M_\alpha$ \\ 
		$\delta \dot{\alpha}_q$ &  Pitch rate to the angle of attack rate map & 	$\sigma(\dot{\alpha}_q) = 0.25 \dot{\alpha}_q$ \\ 
		$\delta N_\alpha$ &  Normal force gradient & 	$\sigma(N_\alpha) = 0.25 N_\alpha$ \\ 
		$\omega_u$ & Input gain & $\sigma(\omega_u) = 0.2$ \\ \hline
	\end{tabular}
\end{table} 

\begin{figure}[h!]
	\centering
	\includegraphics[width=\textwidth]{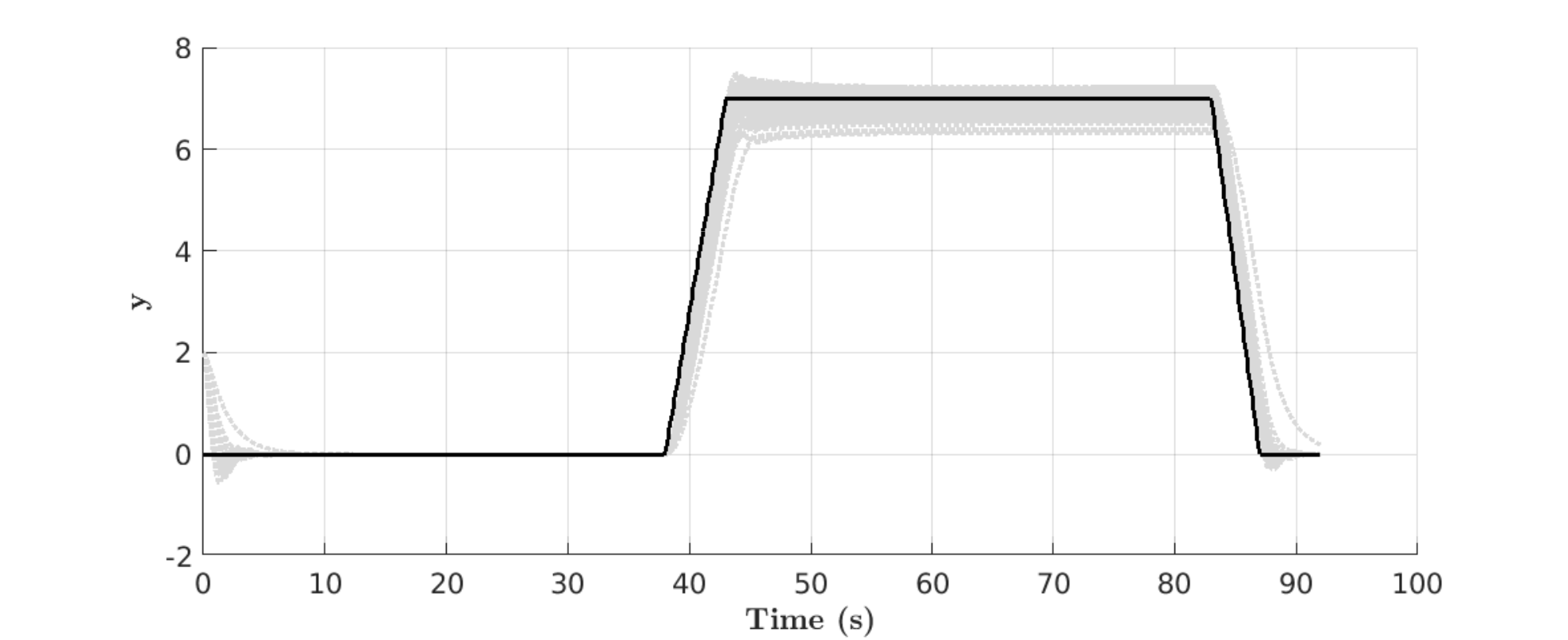}
	\caption{\label{fig:Monte_y_AMPC} Baseline AMPC overlaid Monte Carlo results}
\end{figure}

\begin{figure}[h!]
	\centering
	\includegraphics[width=\textwidth]{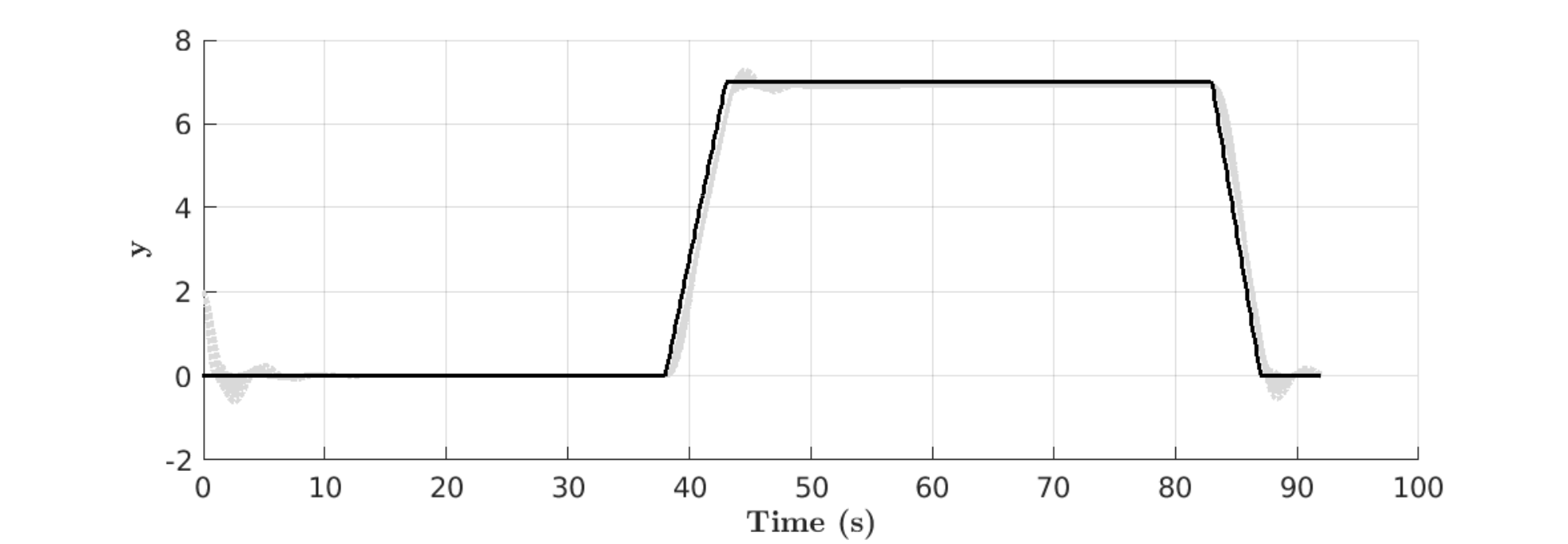}
	\caption{\label{fig:Monte_y_l1} AMPC-$\mathcal{L}_1$ overlaid Monte Carlo results}
\end{figure}

By using the probability distribution from Table \ref{uncertain_params}, one hundred stochastic simulations were computed for both baseline AMPC and AMPC-$\mathcal{L}_1$ controllers. The same uncertainties were used to calculate the responses of both control schemes. From the simulation results overlayed in Figures \ref{fig:Monte_y_AMPC} and \ref{fig:Monte_y_l1}, it is apparent that baseline AMPC has a larger tracking error mean and standard deviation. The means of the tracking error 2-norm for baseline and adaptive AMPC are 97 and 90, respectively, while the standard deviations are 29.1 and 8.0, respectively.

The initial step response is similar for both schemes, but the AMPC-$\mathcal{L}_1$ exhibits consistently superior performance in the ramp response from $t = 40s$ to $t=90 s$. While the baseline AMPC results show a spread of steady-state errors resulting from the parametric uncertainty, the AMPC-$\mathcal{L}_1$ can adapt the control signal to account for the model mismatch and uncertain input gain.


\newpage

\subsection{Robustness Analysis}
\label{sec:robustness}
\paragraph{AMPC Robustness Margins:}
In unconstrained AMPC, an optimal gain matrix is computed according to the Equation \eqref{eqn2.26}. This allows LTI robustness metrics to be applied to the resulting closed-loop system. However, the booster re-entry model is an LTV system. Therefore, LTI models are generated at various operating conditions for the computation of phase and gain margins. The gain and phase margins along the re-entry trajectory are shown in Table \ref{lti_robustness}. 

\begin{table}[h!]
	\centering
	\small
	\caption{\label{lti_robustness} LTI robustness margins for AMPC}
	\begin{tabular}{lllllll}
		\hline
		Mach & Altitude (m) & Phase Margin ($^\circ$) & Gain Margin  \\ \hline
		5 & 26,000 & 125 & $\infty$\\ 
		4 & 25,000 & 126 & $\infty$\\ 
		3 & 23,000 & 125 & $\infty$\\ 
		2 & 17,000 & $\infty$ & $\infty$\\ 
		1 & 13000 & $\infty$ & $\infty$\\ \hline
	\end{tabular}
\end{table} 

\paragraph{Time Delay Margins:}
For adaptive controllers with nonlinear adaptive laws, phase and gain margins are unsuitable as a robustness metric. A commonly used robustness metric for Model Reference Adaptive Controllers (MRAC) is the time delay margin (TDM). The TDM is the time delay required to render the closed-loop system unstable. Although an analytical method for computing TDM has not yet been derived for this particular $\mathcal{L}_1$ adaptive control scheme, the TDM may be computed numerically by gradually increasing the time delay margin until the closed-loop system exhibits signs of instability.

The TDM values are shown in Table \ref{tdm_robustness}. It is clear that the $\mathcal{L}_1$ adaptive augmentation has severely reduced the TDM of the standard AMPC. This is an expected result, due to the trade-offs associated with robustness and performance. For $\mathcal{L}_1$  adaptive control architectures, this trade-off is managed by choosing the cut-off frequency associated with the low-pass filter $C(s)$. Choosing a lower cut-off frequency reduces performance but increases TDM, which will remain bounded regardless of the adaptation rate, allowing for adaptive updates at arbitrarily high frequencies. 

\begin{table}[h!]
	\centering
	\small
	\caption{\label{tdm_robustness} Time delay margins for AMPC and AMPC-$\mathcal{L}_1$ controllers}
	\begin{tabular}{llll}
		\hline
		Mach & Altitude (m) & AMPC (ms) & AMPC-$\mathcal{L}_1$ (ms)\\ \hline
		5 & 26,000 & 300 & 67\\ 
		4 & 25,000 & 300 & 67\\ 
		3 & 23,000 & 310 & 67\\
		2 & 17,000 & 280 & 64\\ 
		1 & 13000 & 287 & 65 \\ \hline
	\end{tabular}
\end{table} 

\subsection{Computational Performance}
One of the claims in this note is that AMPC and AMPC-$\mathcal{L}_1$ using single-point prediction are more computationally efficient than conventional constrained MPC solved using quadratic programming. To verify the computational performance of baseline AMPC and AMPC-$\mathcal{L}_1$, the computational load of these controllers are benchmarked against conventional MPC. All computations have been performed using MATLAB on an Intel Core i7-8550U CPU @ 1.80GHz with 8GB RAM.

For conventional MPC, an interior point method was used to solve the quadratic program. The state transition matrix was computed using a 5th order Taylor series expansion. For AMPC, the eigendecomposition method was used to evaluate the state transition matrix. A single AMPC control update includes the computation of Equations \eqref{eqn3.3}, \eqref{eqn2.26}, \eqref{eq:F}, and \eqref{eq:G}. For AMPC-$\mathcal{L}_1$, the state predictor, adaptive law, and control laws are computed in addition to the calculations required by baseline AMPC. These include Equations \eqref{eq:state_predictor}, \eqref{eq:adaptive_law}-\eqref{eq:adaptive_law5}, and \eqref{eq:control_law}.

The required computation times for each controller to compute a single control update are tabulated in Table \ref{hardware_timing}. Compared to conventional MPC with 10 prediction points, performing a single AMPC update is approximately 15 times faster. With an $\mathcal{L}_1$  augmentation, the computational load is more than baseline AMPC, but is still approximately 10 times faster than conventional MPC with 10 prediction points. 
\begin{table}[h!]
	\centering
	\small
	\caption{\label{hardware_timing} Comparison of computational load}
	\begin{tabular}{lll}
		\hline
		Control scheme & Time for single control update (s)  & Prediction points\\ \hline
		MPC & 0.0208 & 10\\
		MPC & 0.0073 & 5\\
		AMPC & 0.0013 & 1\\ 
		AMPC-$\mathcal{L}_1$ & 0.0022  & 1\\ \hline
	\end{tabular}
\end{table} 

\section{Conclusions}
\label{sec:conclusions}
%This note presents a novel adaptive and efficient model predictive control scheme suitable for use at high update rates. This was done by augmenting the AMPC with $\mathcal{L}_1$ adaptive control. It is shown that the unconstrained AMPC optimal control law may be used to cast the open-loop state space model into partial closed-loop form for adaptive control. These control methodologies are applied to an LTV flyback booster re-entry model as an example application. AMPC-$\mathcal{L}_1$  does not require pre-computed solutions as explicit methods do, but rather relies on accurate computation of the state transition matrix via an eigendecomposition to utilize only a single prediction point. A comparative study between the baseline AMPC and AMPC-$\mathcal{L}_1$ augmented controllers showed that the AMPC-$\mathcal{L}_1$ yields consistently better performance than the baseline in the presence of reduced input gain, model mismatch, and disturbances, based on the tracking error 2-norm. In the simulation test cases, AMPC-$\mathcal{L}_1$ exhibited desirable suppression of oscillations and steady state error. However, the superior performance comes at the cost of a drastically reduced time delay margin. The trade-off between performance and robustness is managed by selecting the cut-off frequency associated with the $\mathcal{L}_1$ adaptive control low pass filter. 

In this note, a novel adaptive and efficient model predictive control scheme for use at high update rates has been developed and applied to an LTV fly back booster re-entry model in simulation studies. The AMPC relies on the accurate computation of the state transition matrix via an eigendecomposition to utilize only a single prediction point and does not require any pre-computed solutions. However, the control performance of baseline AMPC degrades in the presence of model mismatch and disturbances. A comparative study between the baseline AMPC and AMPC-$\mathcal{L}_1$ controllers has shown that the AMPC-$\mathcal{L}_1$ yields consistently better performance than the baseline in the presence of reduced input gain, model mismatch, and disturbances, based on the tracking error 2-norm. Moreover, the AMPC-$\mathcal{L}_1$ has exhibited desirable suppression of oscillations and eliminated steady-state error with the cost of a drastically reduced time delay margin. The trade-off between performance and robustness is managed by selecting the cut-off frequency associated with the $\mathcal{L}_1$ adaptive control low-pass filter. Future work would include hardware-in-the-loop experiments to verify that the proposed controller is computationally feasible on existing embedded systems.

\section*{Appendix}
\label{sec:model}
\subsection{Booster Re-entry Nominal Trajectory}
\begin{figure}[h!]
	\centering
	\includegraphics[width=\textwidth]{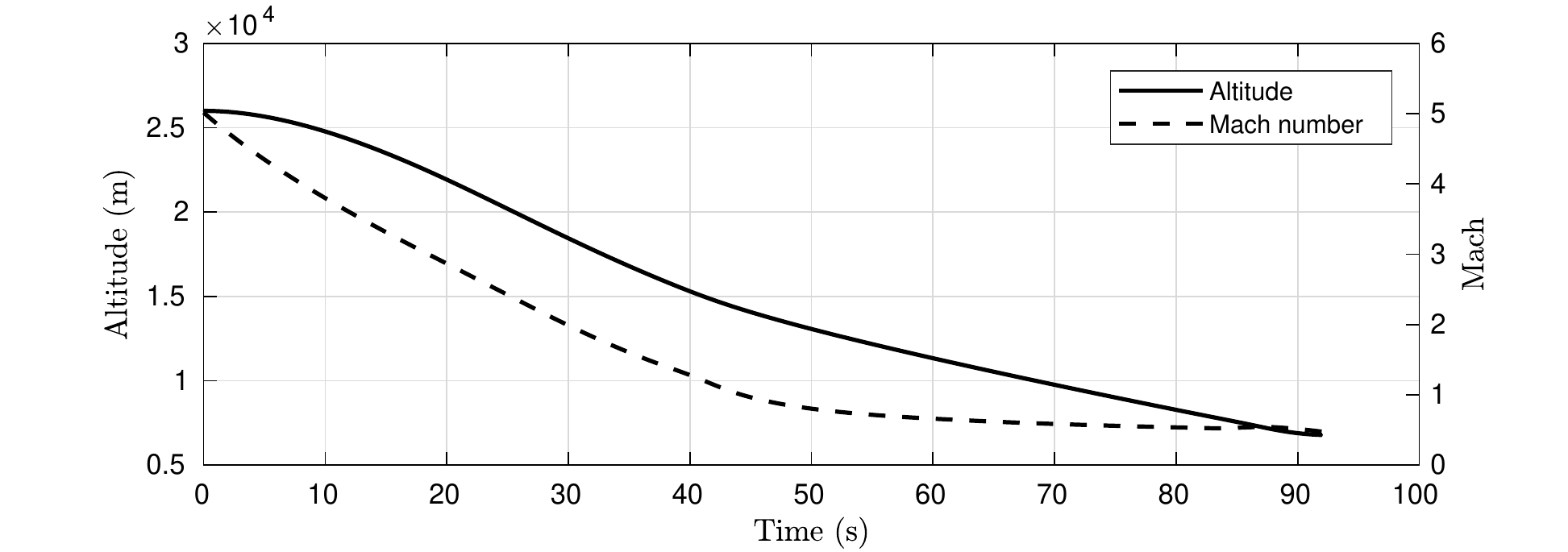}
	\caption{\label{fig:nominal_trajectory} Nominal booster re-entry trajectory}
\end{figure}

\begin{figure}[h!]
	\centering
	\includegraphics[width=\textwidth]{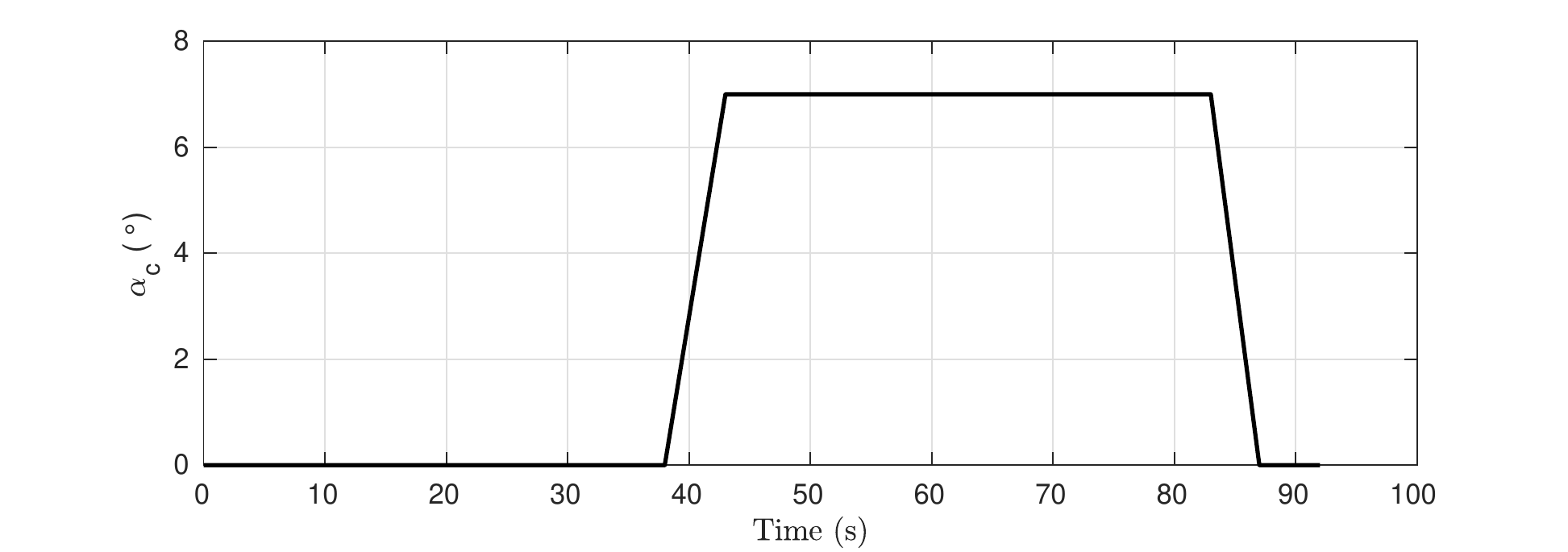}
	\caption{\label{fig:alpha_schedule} Angle of attack command schedule for pull up maneuver}
\end{figure}

\begin{figure}[h!]
	\centering
	\includegraphics[width=\textwidth]{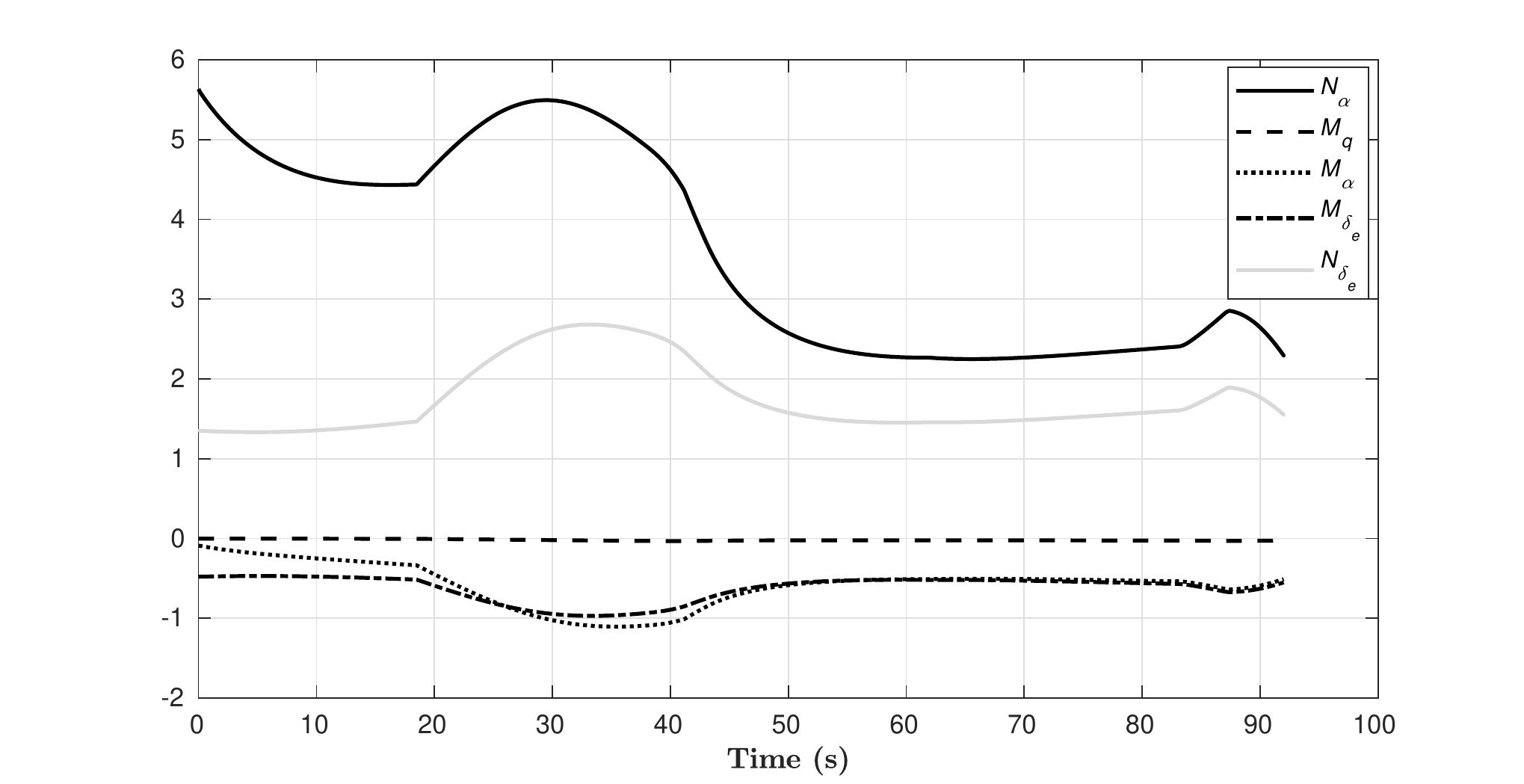}
	\caption{\label{fig:aero_derivs} Time evolution of longitudinal aerodynamic derivatives used for the LTV model}
\end{figure}

In this study, an LTV model is used to design the baseline and augmented AMPC controllers. To derive the LTV model, a nominal trajectory of the booster is utilized to generate the time evolution of the aerodynamic derivatives and velocity, which form the basis of the LTV state-space model described in the subsequent section. The booster model used to generate the LTV model is a reusable launch vehicle first stage concept with tail-fins for re-entry control \cite{Chai2018}. The re-entry trajectory occurs following stage separation (see Figure \ref{fig:nominal_trajectory}). The re-entry starts at Mach 5, 26000 km altitude with a ballistic phase, regulating $\alpha$. A pull up maneuver is initiated at $t=38$ seconds in the form of a ramp $\alpha$ command. The maneuver ends at $t=83$ seconds with a down ramp command, shown in Figure \ref{fig:alpha_schedule}. The time-varying aerodynamic derivatives to be used in the LTV model are shown in Figure \ref{fig:aero_derivs}.


\subsection{Linearized Longitudinal Dynamics}
An angle of attack controller is required to track the commands shown in Figure \ref{fig:alpha_schedule}. This section presents a brief description of longitudinal dynamics along with an LTV model, which is used for control synthesis. The longitudinal motion of aerodynamic vehicles consists of two modes: 1) short period oscillation, which involves high-frequency oscillation, and 2) phugoid, the low-frequency mode. When the phugoid mode is neglected, the aerodynamic derivatives that govern the short period longitudinal motion are pitch stiffness $C_{m_\alpha}$, pitch damping $C_{m_q}$, and normal force gradient $C_{N_\alpha}$. The control authority of the elevators is captured in the derivatives $C_{m_{\delta_e}}$ and $C_{N_{\delta_e}}$. The linearized, short period angle of attack dynamics are
\begin{align}
\label{eq:long_dynamics}
\dot{x}(t) = A(t)x(t) + B_m(t) u(t) &= 
\begin{bmatrix}
\dot{q}(t) \\ \dot{ \alpha}(t)
\end{bmatrix}
 = 
 \begin{bmatrix}
 M_q(t)	& M_{\alpha}(t) \\
 \dot{\alpha}_q & -N_{\alpha}(t)/V(t)
 \end{bmatrix}
 \begin{bmatrix}
 q(t) \\ \alpha(t)
 \end{bmatrix}
 + 
 \begin{bmatrix}
M_{\delta_e}(t) \\ \frac{-N_{\delta_e}(t)}{V(t)}
 \end{bmatrix}
 u(t) \\
 y(t) = C x(t) &= 
 \begin{bmatrix}
 0 & 1
 \end{bmatrix}
 \begin{bmatrix}
 q(t) \\ \alpha(t)
 \end{bmatrix},
\end{align}
where 
\[
M_\alpha = \frac{\bar{q} \bar{S} c}{I_2}C_{m_\alpha},\quad N_\alpha = \frac{\bar{q}\bar{S}}{\bar{m}}C_{N_\alpha},\quad  M_q = \frac{\bar{q}\bar{S}c^2}{2I_2 V} C_{m_q},\quad M_{\delta_e} = \frac{\bar{q}\bar{S}c}{I_2}C_{m_{\delta_e}},\;  N_{\delta_e} = \frac{\bar{q}\bar{S}}{\bar{m}}C_{N_{\delta_e}}. 
\]
The parameter $\dot{\alpha}_q$ is the pitch rate to the angle of attack map, which is assumed to be one-to-one nominally. Aerospace vehicles have highly nonlinear and time-varying plant parameters. Therefore, an LTV model is used for simulation in this study, where parameters of $A(t)$ and $B_m(t)$ are scheduled with respect to time according to the nominal trajectory described in the previous section. However, successive linearization is used with an LTI model for the MPC online optimal control computation. Therefore, the plant parameters are updated according to the current flight condition.

\section*{Acknowledgements}
The first author was a recipient of the Australian Government Research Training Program stipend while undertaking the research activities that culminated in the publication of this note. The authors acknowledge Michael Smart for his ideas behind the fly back booster model.

\bibliography{refs}

\end{document}